\documentclass[aps,prl,twocolumn,amsmath,amssymb,floatfix,superscriptaddress,10pt]{revtex4-2}
\usepackage{color}
\usepackage{amsmath}
\usepackage{amsfonts}
\usepackage{amssymb}
\usepackage{url}

\usepackage{graphicx}
\usepackage{esvect}
\usepackage[normalem]{ulem}
\usepackage[inline]{enumitem}
\usepackage{xcolor}
\usepackage{dsfont}
\usepackage{bm}
\usepackage{array}
\newcolumntype{C}[1]{>{\centering\let\newline\\\arraybackslash\hspace{0pt}}m{#1}}
\DeclareMathAlphabet\mathbfcal{OMS}{cmsy}{b}{n}

\newcommand{\blue}[1]{{\color{blue}#1}}

{
 \definecolor{BLACK}{gray}{0}
 \definecolor{WHITE}{gray}{1}
 \definecolor{RED}{rgb}{1,0,0}
 \definecolor{GREEN}{rgb}{0,1,0}
 \definecolor{BLUE}{rgb}{0,0,1}
 \definecolor{CYAN}{cmyk}{1,0,0,0}
 \definecolor{MAGENTA}{cmyk}{0,1,0,0}
 \definecolor{YELLOW}{cmyk}{0,0,1,0}
}

\usepackage{hyperref}
\hypersetup{colorlinks=true,
            citecolor=blue,
            filecolor=blue,
            linkcolor=blue,
            urlcolor=blue}

\begin{document} 

\title{Local spin magnetization in itinerant non-collinear magnets:  \emph{The local spin Berry curvature}}
\author{Doru Sticlet}
\email{doru.sticlet@itim-cj.ro}
\affiliation{National Institute for R\&D of Isotopic and Molecular Technologies, 67-103 Donat, 400293 Cluj-Napoca, Romania}
\author{Fr\'ed\'eric Pi\'echon}
\email{frederic.piechon@universite-paris-saclay.fr}
\affiliation{Laboratoire de Physique des Solides, Universit\'e Paris-Saclay, CNRS, 91405 Orsay, France}

\begin{abstract}
    Conventionally, the local spin magnetization in itinerant magnets is determined from the equilibrium local spin density. 
    Here, we propose a thermodynamic approach in which the local spin magnetization is defined from the response of the system to an infinitesimal external magnetic field.
    The predictions of the two theories are identical for collinear magnets, but differ qualitatively and quantitatively for non-collinear magnets. 
    In the present thermodynamic approach, the spin coherences determine an alternative distribution of local spin magnetization due to the field-induced deformation of the energy eigenstates.
    This effect is captured by a Berry-curvature-like contribution reminiscent of orbital magnetization and has several distinct observable consequences. 
    We explore the differences between the conventional and thermodynamic approaches in several test cases.
\end{abstract}
\maketitle

\blue{Introduction.---}Non-collinear magnetism emerges when competing exchange interactions, spin-orbit coupling, 
or geometric frustration prevent spins from aligning along a common axis~\cite{Dzyaloshinskii1958,Moriya1960,Lacroix2011}. 
The resulting magnetic order parameter exhibits a spatially varying orientation, 
giving rise to spin textures ranging from long-wavelength spin helices and spirals to skyrmions~\cite{Bak1980,Muehlbauer2009,Nagaosa2013,Fert2017,Garst2017,Bogdanov2020,Goebel2021}.
 Spin textures play an important role for the transport properties of itinerant electron systems~\cite{Moriya1985,Kuebler2021}. The spatial variation of the magnetization acts as an emerging gauge field, generating  a Berry curvature that directly influences electronic motion~\cite{Xiao2010}, and leads to unconventional transport phenomena, including anomalous Hall effects without net magnetization and current-induced spin torques~\cite{Nagaosa2010,Manchon2019,Chang2023}.

Beyond transport properties, non-collinearity fundamentally alters the nature of the electronic eigenstates. 
In a collinear magnet, energy eigenstates can be chosen as eigenstates of a global spin projection axis, while in non-collinear spin texture, the energy eigenstates are no longer spin eigenstates and therefore the spin operator acquires finite off-diagonal matrix elements in the energy eigenbasis, i.e.,~spin coherences.
The latter are central in understanding the electronic response of non-collinear magnets, yet their role in determining the local spin magnetization is not fully understood.
The conventional theory sees the local magnetization in the context of an  equilibrium spin density~\cite{Barth1972,Liechtenstein1987,Hobbs2000}. 
In contrast, and maybe more consequential for experiments, 
we develop here a thermodynamic theory for the local spin magnetization as determined by response to an external magnetic field and show that it entails novel qualitative features.

This problem is timely in the context of recent experimental and theoretical advances.
Experimental progress has allowed real-space imaging of magnetic texture at the atomic scale~\cite{Wiesendanger2009,Huang2023}.
Through techniques such as spin-polarized scanning tunneling microscopy (SP-STM)~\cite{Heinze2000,Wortmann2001,Kubetzka2005,Heinze2006,Bode2007,Ferriani2008,Gao2008,Heinze2011}, non-magnetic STM~\cite{VonBergmann2012}, and magnetic exchange force microscopy~\cite{Schmidt2009,Grenz2017,Hauptmann2020}, researchers have been able both to investigate the local orientation of magnetic moments at nanometer scales, and to manipulate and design non-collinear spin textures~\cite{Guo2026}.
Theoretically, the thermodynamic approach has proved fruitful in constructing the modern theory of orbital magnetization~\cite{Thonhauser2005,Xiao2005,Ceresoli2006,Shi2007,Raoux2015}, which has been extended to provide a real-space description of orbital magnetism~\cite{Bianco2011,Bianco2013,Saati2025}.
Building upon these advances, the present theory provides a unified framework for local orbital and spin magnetization.

Conventionally, in a mean-field-like treatment of multiband electronic systems, the local spin magnetization is the spin magnetic moment projection of occupied quasiparticle states on a given atomic site~\cite{Barth1972,Liechtenstein1987,Hobbs2000,Szilva2023}:
\begin{align}
\bm M_{\text{spin}}(\bm r,\mu) = -\mu_B \sum_n  f(\varepsilon_{n}) 
 \langle n | \bm \sigma_{\bm r} | n\rangle
\label{mspinsloc1}
\end{align} 
where $\bm \sigma_{\bm r}\equiv |\bm r \rangle \langle \bm r| \otimes \bm  \sigma$ is the spin operator projected at position $\bm r$, $|n \rangle$ is the quasiparticle eigenstate of energy $\varepsilon_n$, and $f(\varepsilon) = 1/(1+e^{\beta(\varepsilon-\mu)})$ is the Fermi function, with
$\mu$ the chemical potential, $\beta = 1/(k_B T)$ the inverse temperature, and $\mu_B$ the Bohr magneton.
The thermodynamic approach developed in the following determines an alternative expression for the local spin magnetization:
\begin{align}
  \widetilde{\bm M}_{\text{spin}}(\bm r) = -\mu_B 
\sum_n  [f(\varepsilon_{n}) \bm m_{n}(\bm r)+F(\varepsilon_{n})\bm \Omega_{n}(\bm r)],
\label{mspin1}
\end{align}
with 
\begin{align}
\bm m_{n}(\bm r)&=|\langle \bm r|  n\rangle|^2 \langle n|\bm \sigma |n \rangle,\label{mspin_diag}\\ 
\bm \Omega_{n}(\bm r)&=-2\sum_{m\ne n}\textrm{Re} \dfrac{\langle \bm r|  n\rangle
\langle m |\bm r\rangle \langle n|\bm \sigma |m \rangle}{\varepsilon_{n } -\varepsilon_{m }}, 
\label{mspin2}
\end{align}
and $F(\varepsilon) = k_B T \ln\left(1 + e^{-\beta(\varepsilon-\mu)}\right)$.
We will show that the first term~\eqref{mspin_diag}, corresponding the diagonal matrix element of the spin operator, is also present in Eq.~\eqref{mspinsloc1}, while the second term~\eqref{mspin2} is a novel contribution which entails unexpected effects, such as a linear variation of the local spin magnetization in a spectral gap.
Moreover, in Eq.~\eqref{mspin1}, the diagonal and off-diagonal contributions support different physical pictures. 
More specifically, to linear order in an external magnetic field, the diagonal local-moment contribution~\eqref{mspin_diag} yields the field-induced Zeeman shift of the energy levels, whereas the off-diagonal one~\eqref{mspin2} expresses the field-induced deformation of the eigenstates.
The latter~\eqref{mspin2} plays the role of a \emph{local spin Berry curvature} similar to the Berry-curvature term in the local orbital magnetization recently established in Ref.~\cite{Saati2025}.
Remarkably, the thermodynamic theory provides a common formalism for both spin and orbital local magnetization, since substituting the spin magnetic moment $-\mu_B \bm \sigma$ by the orbital magnetic moment  $\bm L=-\frac{e}{2} \bm r\times \bm v$ in Eqs.~(\ref{mspin1})--(\ref{mspin2}) recovers exactly the local orbital magnetization expression~\cite{Saati2025}.

\blue{Conventional theory of local spin magnetization.---}To set the stage, let us briefly revisit the conventional decomposition of local spin magnetization Eq.~\eqref{mspinsloc1} into a diagonal (d) and an off-diagonal (od) contribution,
\begin{equation}
\langle n |\bm \sigma_{\bm r} | n \rangle=\bm m_n^{\textrm{d}}(\bm r)+\bm m_n^{\textrm{od}}(\bm r)
\end{equation} 
with
\begin{align}
\bm m_n^{\textrm{d}}(\bm r)&=|\langle n|\bm r\rangle|^2 \langle n | \bm \sigma | n\rangle,\label{eq:m_r_intra}\\
\bm m_n^{\textrm{od}}(\bm r)&=\sum_{m \ne n} \langle n |\bm r\rangle\langle \bm r |m \rangle  \langle m | \bm \sigma | n \rangle.\label{eq:m_r_inter}
\end{align} 
For a collinear magnetic system, $\bm m_n^{\rm od}(r)$, which involves the off-diagonal spin coherences, vanishes since energy eigenstates are also spin eigenstates. 
For a non-collinear magnetic texture, the off-diagonal term survives, and the total local spin magnetization is generally resolved into a sum of diagonal and off-diagonal contributions,
\begin{equation}
\bm M_{\rm spin}^{\rm d/od}(\bm r) = -\mu_B
\sum_n f(\varepsilon_n)\bm m_n^{\rm d/od}(\bm r).
\label{intra_inter}
\end{equation}
An immediate consequence is that the diagonal terms are identical in both conventional~\eqref{eq:m_r_intra} and thermodynamic approaches~\eqref{mspin_diag}.
To better understand the effect of the off-diagonal terms, we next develop the thermodynamic theory in more detail.

% In this work, we  present a microscopic thermodynamic approach that provides an alternative form for the local spin magnetization of an itinerant electron gas in a non-collinear mean-field spin texture. As detailed below, our novel expressions~\eqref{mspin1}--\eqref{mspin2} 
% show that  while the diagonal contribution is identical to 
% $\bm M^{\rm d}_{\rm spin}(\bm r)$,
% the off-diagonal contribution differs qualitatively and quantitatively from 
% $\bm M^{\rm od}_{\rm spin}(\bm r)$.
% Moreover,  
% We compare our formulation with the conventional approach in a few examples. 

\blue{Thermodynamic theory of local spin magnetization.---}The local magnetization (orbital and spin) is obtained thermodynamically by defining it as the zero-field limit of the first-order derivative of the local, field-dependent grand potential,
$ \bm M(\bm r,\mu) = - \partial_{\bm B} \Xi|_{\bm B=0}$,
with $\Xi(\bm r, \mu, \bm B) = -\int \mathrm{d}\varepsilon \, F(\varepsilon) \, \rho(\bm r,\varepsilon, \bm B)$
with $F(\varepsilon)$ defined above. 
In this expression, $\rho(\bm r,\varepsilon, \bm B)=-\frac{\mathrm{Im}}{\pi}  \langle \bm r|  \mathcal{G}(\varepsilon,{\bm B}) | \bm r \rangle$ is the field-dependent local density of states, which is expressed in terms of the field-dependent single particle Green's function of the system $\mathcal{G}(\varepsilon, \bm B)$. 
To obtain more specifically the spin magnetization, we consider only the Zeeman coupling to the magnetic field. 
Therefore, to linear order in the magnetic field, it follows that
$\mathcal{G}(\varepsilon, \bm B)=\mathcal{G}(\varepsilon) +\mu_B \mathcal{G}(\varepsilon)(\bm B \cdot \bm \sigma) \mathcal{G}(\varepsilon)+ {\cal O}(\bm B^2)$
in terms of the single-particle zero-field Green's function $\mathcal{G}(\varepsilon)=(\varepsilon -{\cal H})^{-1}$.
Using this identity yields the following alternative expression for the local spin magnetization:
\begin{align}
    \label{eq:M_th}
    \widetilde{\bm M}_{\text{spin}}(\bm r) = -\mu_B \int_{-\infty}^{+\infty}\mathrm{d}\varepsilon F(\varepsilon)
    \frac{\mathrm{Im}}{\pi}  \langle \bm r|  \mathcal{G} (\varepsilon)\bm \sigma \mathcal{G}(\varepsilon)| \bm r \rangle.
\end{align}
Employing $\mathcal{G}(\varepsilon)=\sum_n \frac{|n\rangle\langle n|}{\varepsilon-\varepsilon_n}$ in Eq.~\eqref{eq:M_th}, finally determines the novel local spin magnetization in Eqs.~\eqref{mspin1}--\eqref{mspin2}.

While the diagonal contributions are identical in both theories, the off-diagonal contribution $\bm \Omega_{n}(\bm r)$ is quite different from $\bm m_n^{\textrm{od}}(\bm r)$ \eqref{eq:m_r_inter}, leading to
\begin{equation}
\widetilde{\bm M}^{\rm od}_{\rm spin}(\bm r) \equiv -\mu_B\sum_n F(\varepsilon_n) \bm \Omega_n(\bm r) \neq \bm M_{\rm spin}^{\rm od}(\bm r),
\end{equation}
and, consequently, a difference in the total local spin magnetization 
$\widetilde{\bm M}_{\rm spin}(\bm r)\neq \bm M_{\rm spin}(\bm r)$.

To better understand the novel qualitative feature contained in formulas \eqref{mspin1}--\eqref{mspin2}, it is instructive to consider the zero-temperature limit where $f(\varepsilon)=\Theta(\mu-\varepsilon)$ and $F(\varepsilon)=(\mu-\varepsilon)f(\varepsilon)$, with $\Theta$ the Heaviside step function. 
The local spin magnetization rewrites as
\begin{align}
    \widetilde{\bm M}_{\text{spin}}(\bm r) =-\mu_B 
\sum_n  f(\varepsilon_{n}) [\bm {\mathcal M}_n(\bm r)+ \mu \bm \Omega_{n}(\bm r)],
\label{mspin3}
\end{align}
with $\bm {\mathcal M}_n(\bm r)=\bm m_{n}(\bm r)-\varepsilon_n \bm \Omega_{n}(\bm r)$.
The second term of \eqref{mspin3} is the main difference between our thermodynamic formalism and the conventional approach. 
Qualitatively, it implies that the local spin magnetization might vary linearly in a gap as a function of the chemical potential. 
This feature is reminiscent of the (local) Berry-curvature contribution in the (local) orbital magnetization~\cite{Saati2025} which is obtained by substituting the spin magnetic moment $-\mu_B\bm \sigma$ in Eq.~\eqref{mspin2} by the orbital moment operator $\bm L=-\frac{e}{2} \bm r \times \bm v$. 
Therefore, it strongly suggests an interpretation of the term $\bm \Omega_{n}(\bm r)$ as an effective \emph{local spin Berry curvature}. 
Furthermore, it is also straightforward to check that the expressions \eqref{mspin1}--\eqref{mspin2} correspond to substituting in the grand potential an effective local field-dependent density given by
\begin{align}
\rho(\bm r,\varepsilon, \bm B)&=\sum_n (|\langle \bm r | n\rangle|^2-\mu_B \bm \Omega_{n}(\bm r)\cdot \bm B)\notag\\
&\quad\times \delta (\varepsilon-\varepsilon_n-\mu_B \bm m_n \cdot \bm B),
\label{ldos}
\end{align}
where $ \bm m_n =\langle  n|\bm\sigma|n\rangle$.
From Eq.~\eqref{ldos}, we conclude that, in \eqref{mspin1}, the diagonal contribution $\bm m_{n}(\bm r)$ may be understood as due to a field-induced Zeeman shift of the energy bands, whereas the off-diagonal contribution associated with the \emph{local spin Berry curvature} $\bm \Omega_{n}(\bm r)$ corresponds instead to an effective field-induced local deformation of the wavefunction.
The expressions \eqref{mspin1}--\eqref{mspin2}, together with the relation~\eqref{ldos}, constitute the main result of this work. 

Another important property of our thermodynamic approach is that, in contrast to the conventional expression~\eqref{mspinsloc1}, the Eqs.~\eqref{mspin1}--\eqref{mspin2} necessarily verify the thermodynamic identity $\partial \bm M (\bm r)/\partial \mu=\partial n(\bm r)/\partial \bm B$, where $n(\bm r,\mu,\bm B)=-\partial_\mu \Xi$ is the local field-dependent particle density. 
This identity implies that the local (spin) magnetization might be experimentally measured with a local (non-magnetic) STM probe by examining  the field dependence of the local particle density. 

In the following, we consider a few examples to provide a more quantitative and contrasting view of local spin magnetization in the conventional~\eqref{mspinsloc1} and thermodynamic~\eqref{mspin1}--\eqref{mspin2} theories. 

\begin{figure}[t]
    \centering
    \includegraphics[width=\columnwidth]{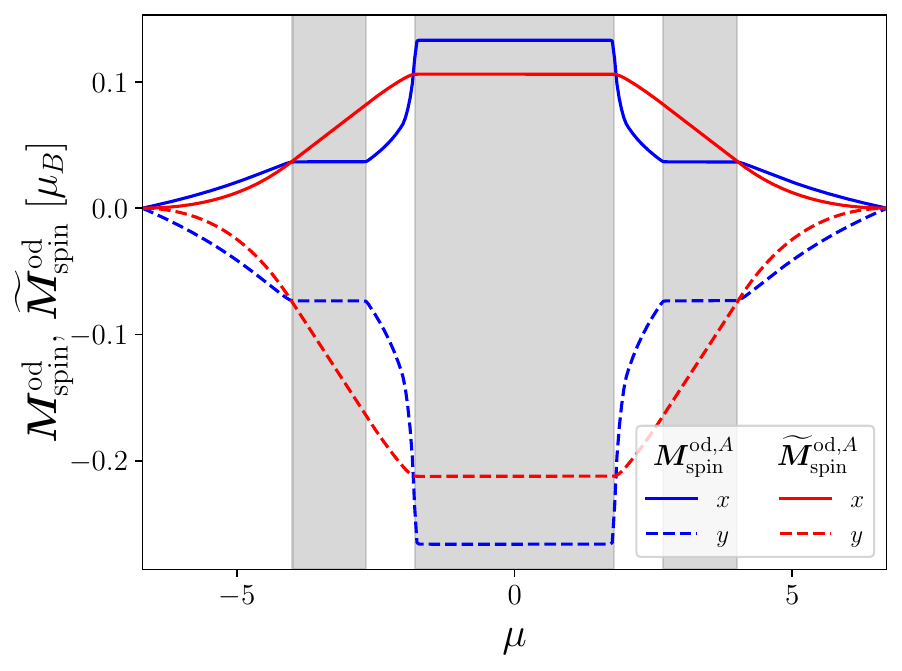}
    \caption{Off-diagonal (interband) local spin magnetization as a function of the chemical potential in the conventional theory $\bm M^{\rm od}_{\rm spin}$ (blue) and the thermodynamic one $\widetilde{\bm M}^{\rm od}_{\rm spin}$ (red) for the ferrimagnet described by Eq.~\eqref{eq:ham_sq}. The plot shows only the $x$ (continuous line) and $y$ (dashed line) components of the off-diagonal magnetization on the $A$ sublattice (for the $B$ sublattice it takes the opposite values). 
    The results are in the strong-coupling regime with effective exchange fields $\bm h_A=(4,0,0)t$ and $\bm h_B=(0,2,0)t$ for a temperature $T=0.01t$. The gray regions mark the energy gaps.
    }
    \label{fig:comp_sq}
\end{figure}

\blue{Four-band non-collinear ferrimagnet.---}As a first example, we consider a four-band tight-binding model of a non-collinear ferrimagnet on a square lattice with in-plane exchange fields $\bm h_A$ and $\bm h_B$ on $A$ and $B$ sublattices, respectively, a uniform external magnetic field $\bm B$,
and a spin-independent nearest-neighbor hopping of amplitude $t$ between the two sublattices. 
The Bloch Hamiltonian for the system reads
\begin{align}
H_{\bm k}= (\bm h_++\bm B) \cdot \bm \sigma + (\bm h_-\cdot \bm \sigma) \tau_z +g_{\bm k} \tau_x,
\label{eq:ham_sq}
\end{align}
with $\bm h_\pm=(\bm h_A \pm \bm h_B)/2$, $g_{\bm k}=2 t (\cos k_x+\cos k_y)$,
and where $\bm \sigma$ and $\bm \tau$ are the spin and sublattice pseudospin vectors of Pauli matrices.
The Hamiltonian $H_{\bm k}$ is a $4\times4$ matrix where $\sigma_\alpha\equiv\mathds{1}\otimes \sigma_\alpha$ and $\tau_\alpha\equiv\tau_\alpha\otimes \mathds{1}$, with $2\times2$ identity matrix $\mathds{1}$, and $\alpha\in\{x,y,z\}$.
To further simplify the expressions, we rewrite $\bm h_{+}^{\bm B}=\bm h_++\bm B$.
Note that from now on we work in units where $\mu_B=1$, so that the quantities $\bm h_{A,B},\bm h_\pm,\bm B$ are effectively homogeneous to an energy, which itself is computed in units of hopping $t$.

The main interest of this model is that its physical properties are obtained analytically to any order in the external magnetic field $\bm B$. 
In this Bloch-band picture, the diagonal and off-diagonal contributions defined in Eqs.~\eqref{eq:m_r_intra}, \eqref{eq:m_r_inter}, and \eqref{mspin2} now correspond to one intraband and two interband contributions, ${\bm m}^{\rm d, \alpha}_{n \bm k}$, ${\bm m}^{\rm od, \alpha}_{n \bm k}$, and ${\bm \Omega}_{n \bm k}^\alpha$, respectively, with $n$ the band index, $\bm k$ the momentum, and $\alpha=A,B$ the sublattice index.
The local spin magnetization on each sublattice $\bm r_\alpha$ reads
\begin{align}
{\bm M}_{\text{spin}}^\alpha &=-
\sum_n  \langle f(\varepsilon_{n \bm k}) 
[{\bm m}^{\rm d, \alpha}_{n \bm k}
+ \bm m^{\rm od, \alpha}_{n \bm k}]\rangle_{\bm k},\\
\widetilde{\bm M}_{\text{spin}}^\alpha &=-
\sum_n \langle f(\varepsilon_{n \bm k}) \bm  m^{\rm d,\alpha}_{n \bm k}+ F(\varepsilon_{n \bm k})  
\bm \Omega_{n \bm k}^\alpha\rangle_{\bm k},
\end{align}
with Brillouin zone averaging $\langle\ldots\rangle_{\bm k}\equiv\int d^d\bm k/(2\pi)^d(\ldots)$.
The exact expressions for the three contributions follow after straightforward calculations using explicit projectors~\cite{Graf2021} onto the bands of the Hamiltonian~\eqref{eq:ham_sq},  
\begin{align}\label{eq:omegak}
{\bm m}^{\rm d,\alpha}_{n \bm k }=&\frac{1}{2}(1+s_\alpha \tau_n v_{\bm k})\frac{1}{\varepsilon_{n \bm k}}((1+\tau_n u_{\bm k}) \bm h_+ +\tau_n v_{\bm k} \ \bm h_-),\notag \\
{\bm m}^{\rm od,\alpha}_{n \bm k }=&\frac{s_\alpha }{2}\frac{u_{\bm k}}{\varepsilon_{n \bm k}\Delta_{\bm k}} (\bm h_+ \times \bm h_-)\times \bm h_+,\\
{\bm \Omega}^{\alpha}_{n\bm k}=&\frac{s_\alpha }{2}\tau_n \frac{u_{\bm k}}{\Delta^2_{\bm k}}(\bm h_- \times \bm h_+)\times \bm h_+, \notag
\end{align}
with $s_{\alpha}=+/-$ for the $A/B$ sublattice,  $u_{\bm k}=g^2_{\bm k}/\Delta_{\bm k}$, $v_{\bm k}=(\bm h_- \cdot \bm h_+ )/\Delta_{\bm k}$, and $\varepsilon_{n \bm k}=\sigma_n \sqrt{\Delta_{\bm k}^0+2\tau_n \Delta_{\bm k}}$ with $\Delta_{\bm k}^0=(|\bm h_+ |^2+ |\bm h_- |^2+ g^2_{\bm k})$, $\Delta_{\bm k}=\sqrt{ g^2_{\bm k} |\bm h_+ |^2+ (\bm h_- \cdot \bm h_+ )^2}$, and $\sigma_n=\pm, \tau_n=\pm$ depending on the band index $n$.

Essentially, two main qualitative features can be deduced from Eq.~\eqref{eq:omegak}. 
First, when $(\bm h_+\cdot  \bm h_-)=0$ ($v_{\bm k} =0$), the intraband part is the same on each sublattice and only along $\bm h_+$. 
Second, the interband terms ${\bm m}^{\rm od,\alpha}_{n \bm k }$ and ${\bm \Omega}^{\alpha}_{n \bm k }$ are oriented along the same direction but differ in amplitude.
More quantitatively, Fig.~\ref{fig:comp_sq} shows the off-diagonal (interband) local spin magnetization as a function of the chemical potential in the conventional theory $\bm M^{\rm od}_{\rm spin}$ (blue) and the thermodynamic theory $\widetilde{\bm M}^{\rm od}_{\rm spin}$ (red). 
The plot shows only the two nonzero off-diagonal magnetization components for the $A$ sublattice (for the $B$ sublattice it takes opposite values). 
As expected, the main qualitative difference appears in the energy gaps that separate the four bands. 
While the conventional theory necessarily gives a plateau of magnetization in the gap, the thermodynamic theory allows for a linear variation of the sublattice magnetization that is solely due to the \emph{local spin Berry curvature} and is reminiscent of what is observed for the sublattice orbital magnetization~\cite{Saati2025}.
The Supplemental Material (SM)~\cite{SM} shows that the same phenomenology  is reproduced at weaker magnetic fields with an additional sublattice potential, even in a magnetic dimer model, a zero-dimensional toy-model of Eq.~\eqref{eq:ham_sq}, and in a magnetically gapped dice lattice with distinct exchange fields $\bm h_{A,B,C}$ on each of its three sublattices.
Quite generically, in crystalline systems with arbitrarily many sublattices, conventional and thermodynamic theories predict different distributions for the local spin magnetization among sublattices due to the different interband contributions.
Nevertheless, these differences integrate to zero over a complete unit cell, so that only site-resolved probes may distinguish the two theories.

\begin{figure}[t]
\includegraphics[width=\columnwidth]{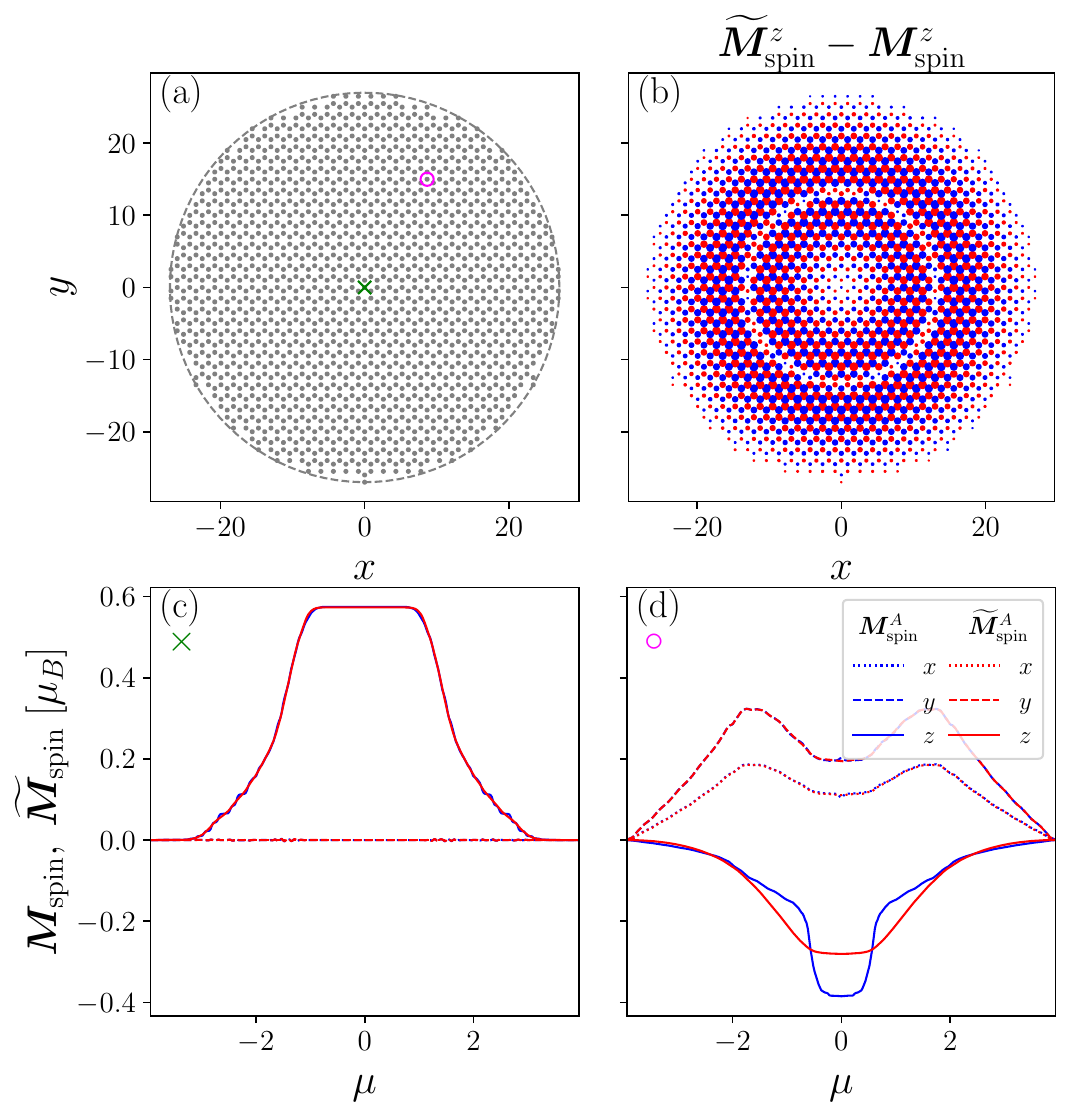}
\caption{(a) The sites of the honeycomb lattice circumscribed by a circle of radius $R=27a$ ($1724$ sites). 
The symbols $\times$ and $\circ$ mark the $A$ sites at which the local spin magnetization is computed for all $\mu$ in panels (c) and (d), respectively.
(b) The difference between the $z$ components of the local spin magnetization in the thermodynamic and conventional theories, for all sites in the lattice and $\mu=0$.
Blue and red colors indicate negative and positive values, respectively, while the size of the markers is proportional to the magnitude of the difference. 
The local spin magnetization as a function of $\mu$ at the central A site (c) and an arbitrary A site (d), respectively. In (d), the difference between the two theories is apparent in the $z$ component.
The legend is shared in (c) and (d), with blue lines denoting $\bm M_{\rm spin}$ and red lines, $\widetilde{\bm M}_{\rm spin}$.
The skyrmion field amplitude is $M=t$ and temperature, $T=0.01t$.
}
\label{fig:skyrmion}
\end{figure}

\blue{Skyrmions on a honeycomb lattice.---}The theory of local spin magnetization is well-suited for applications to real space where nontrivial magnetic textures are present.
In the following, we illustrate quantitative differences between the conventional and thermodynamic theories in the case of opposite-polarity skyrmion textures on a honeycomb lattice.
The tight-binding Hamiltonian for the system reads
\begin{equation}
\label{eq:H_skyrmion}
H = \sum_{i,\alpha}[\bm c^\dag_{\alpha,i} \bm h_\alpha(\bm r_i)\cdot\bm \sigma c_{\alpha,i}]+t\sum_{\langle i,j\rangle} (c_{A,i}^\dag c_{B,j} + \text{h.c.}),
\end{equation}
with creation operators $c_{\alpha,i}^\dag=(c_{\alpha,i,\uparrow}^\dag, c_{\alpha,i,\downarrow}^\dag)$ for sublattice $\alpha=A,B$, $i$ labelling the unit cells, and $\langle \dots\rangle$ denoting summation over nearest-neighbor sites.
There is a finite system centered at the origin, with a circular geometry of radius $R$ [Fig.~\ref{fig:skyrmion}(a)].
The local exchange fields $\bm h_{A,B}(\bm r_i)$ form a skyrmion texture, with the skyrmion field on $A$ and $B$ having opposite polarity,
\begin{equation}
\bm h_\alpha(\bm r) = M\left(\frac {x}{r}\sin\theta_r, \frac {y}{r}\sin\theta_r, s_\alpha\cos\theta_r \right),
\end{equation}
with $M$ the magnitude of the exchange field, $r=|\bm r|$, $\theta_r=\pi r/R$, and $s_{A/B}=\pm 1$.
Thus, the finite lattice fits exactly one skyrmion on each sublattice.
On the A (B) sublattice, the spins point parallel (antiparallel) to the $z$ axis at the skyrmion center and reverse radially their orientation to point antiparallel (parallel) at the lattice edge.

The local spin magnetization in the conventional and thermodynamic theories is computed numerically using Eqs.~\eqref{mspinsloc1} and~\eqref{mspin1}. 
While $x$ and $y$ components of $\bm M_{\rm spin}$ and $\widetilde{\bm M}_{\rm spin}$ are quantitatively similar, marked differences appear in the $z$ component.
Fig.~\ref{fig:skyrmion}(b) shows the difference between $\widetilde{\bm M}^z_{\rm spin}$ and $\bm M^z_{\rm spin}$ at fixed chemical potential on all lattice sites.
Notably, the difference is minimal at $r\simeq 0$, $r\simeq R/2$, and lattice boundary $r\simeq R$, since the exchange fields on neighboring sites are almost collinear there, up to small angular corrections on the order of $a\pi/R$,
[antiparallel at $r\simeq 0$ and $r\simeq R$, and parallel at $r\simeq R/2$].
Away from these limits, the difference becomes quantitatively non-negligible.
Figs.~\ref{fig:skyrmion}(c) and (d) show the full local magnetization for all chemical potentials at $r\simeq 0$ and at some arbitrary point in the lattice, respectively.
The latter case, in Fig.~\ref{fig:skyrmion}(d), illustrates the difference that arises in the $z$ component of magnetization, in contrast to the in-plane components ($x$ and $y$) which display similar features in both theories.
The SM~\cite{SM} includes an additional example in a hexagonal lattice, where the interplay between Rashba spin-orbit coupling and a single skyrmion texture generates a linear variation of magnetization in the gap opened by a staggered onsite potential.

%\subsection{Three sublattices, 6-band models}

\blue{Conclusions.---}We have shown that there are two inequivalent notions of local spin magnetization in itinerant non-collinear magnets.
One is the equilibrium local spin density that is the focus in the conventional theory.
The second one, developed here, is based on the thermodynamic response to a magnetic field. 
In the latter case, the local spin magnetization contains two physically distinct contributions. 
The diagonal term is the familiar local-moment response associated with the Zeeman shift of quasiparticle energies, whereas the off-diagonal one is controlled by spin coherences and describes the eigenstates' field-induced deformation.
Within the thermodynamic formulation, the off-diagonal contribution naturally takes the form of a local spin Berry-curvature term, which has no counterpart in the conventional expression based on the equilibrium local spin density.

The difference between the two theories of the local spin magnetization is not only formal but also leads to qualitatively and quantitatively distinct outcomes in non-collinear magnets.
A most striking prediction of the thermodynamic theory is that of a possible linear variation for the local spin magnetization inside the spectral gaps, as we have illustrated in a square lattice ferrimagnet, and several additional examples in the SM~\cite{SM}.
Quite commonly, the two theories also predict quantitative differences for the distribution of local magnetization, as was shown for a skyrmion textures on a hexagonal lattice.
However, by construction, the off-diagonal corrections to local spin magnetization, either in the form of the conventional off-diagonal components of the projected spin operator or in the novel Berry-curvature-like form predicted in the thermodynamic theory, cancel out in crystalline systems when summing the magnetizations in a complete unit cell, or when averaging over the lattice sites in real-space models involving nontrivial spin textures which break the lattice translation symmetry.
Thus, the discrepancy between them is invisible to bulk probes, which may explain why such a fundamental aspect of local spin magnetization has remained unnoticed so far.
Our results therefore motivate the development and application of local magnetic probes capable of resolving magnetization within a unit cell, such as spin-polarized STM and related spin-resolved spectroscopies~\cite{Wiesendanger2009,Huang2023}.

More broadly, the local spin Berry curvature introduced here is formally analogous to the local orbital magnetization Berry curvature that quantifies the off-diagonal response of the orbital magnetization to an external magnetic field.
Thus, we hope that this common theoretical framework will serve as a useful organizing concept for the experimental investigation and theoretical analysis of the total local magnetization in itinerant non-collinear magnets.

% \begin{acknowledgments}
\blue{Acknowledgments.---}This work was supported by a grant of MCID, CCCDI - UEFISCDI, project numbers PN-IV-P1-PCE-2023-0987, PN-IV-P1-PCE-2023-0159 and PN-IV-P8-8.3-PM-RO-FR-2024-0059, and by the “Nucleu” Program within the PNCDI 2022-2027, Romania, carried out with the support of MEC, project No. 27N/03.01.2023, component project code PN 23 24 01 04.
% \end{acknowledgments}

\bibliography{references}
\end{document}

\title{Supplemental Material for ``Local spin magnetization in itinerant non-collinear magnets: \emph{The local spin Berry curvature}''}
\author{Doru Sticlet}
\email{doru.sticlet@itim-cj.ro}
\affiliation{National Institute for R\&D of Isotopic and Molecular Technologies, 67-103 Donat, 400293 Cluj-Napoca, Romania}
\author{Fr\'ed\'eric Pi\'echon}
\email{frederic.piechon@universite-paris-saclay.fr}
\affiliation{Laboratoire de Physique des Solides, Universit\'e Paris-Saclay, CNRS, 91405 Orsay, France}

\begin{abstract}
The Supplemental Material provides detailed derivation of the results in the main article, together with additional examples that contrast the thermodynamic and conventional theories of local spin magnetization.
Sec.~\ref{sec:4fer} details the analytical derivation of the expression for the local spin magnetization in the four-band non-collinear ferrimagnet on a square lattice, with additional numerical results regarding the distribution of spin magnetization in momentum space and the action of an onsite staggered potential.
Sec.~\ref{sec:dimer} illustrates analytically that the two theories predict distinct local magnetization distributions even in a minimal magnetic dimer model.
Sec.~\ref{sec:dice} numerically investigates a three-sublattice system, the magnetic dice model.
Finally, Sec.~\ref{sec:soc_skyrmion} provides additional numerical results in a real space model on a hexagonal lattice that features a skyrmion field, spin-orbit coupling, and a staggered onsite potential.
\end{abstract}

\maketitle
\section{Four-band non-collinear ferrimagnet}\label{sec:4fer}
\subsection{Derivation of intraband and interband contributions to the local spin magnetization}
In this section, we detail the derivation and the results for the four-band non-collinear ferrimagnet on a square lattice presented in the main text.
The Hamiltonian of the system is given by Eq.~(13) in the main text,
\begin{equation}\label{eq:ham_sq_sm}
H_{\bm k}= (\bm h_++\bm B) \cdot \bm \sigma + (\bm h_-\cdot \bm \sigma) \tau_z +g_{\bm k} \tau_x,
\end{equation}
with the energy band spectrum
\begin{align}
\varepsilon_{n{\bm k}}(\bm B)=\sigma_n \sqrt{\Delta^0_{\bm k}+2 \tau_n \Delta_{\bm k}}
\end{align}
with $(\sigma_n,\tau_n)\equiv \pm$ and
\begin{align}
%\begin{array}{l}
\Delta^0_{\bm k} &=|\bm h^{\bm B}_+ |^2+ |\bm h_- |^2+ g^2_{\bm k},\\
\Delta_{\bm k} &=\sqrt{ g^2_{\bm k} |\bm h^{\bm B}_+ |^2+ (\bm h_- \cdot \bm h^{\bm B}_+ )^2}.
%\end{array}
\end{align}
Denoting $|n\bm k\rangle$ the associated eigenstate, the corresponding eigenprojector $P_{n\bm k}(\bm B)=|n\bm k\rangle \langle n\bm k|$ takes the form~\cite{Graf2021}
\begin{equation}
\label{eq:projectors}
P_{n{\bm k}}=\frac{1}{4}\left(1+\frac{H_{\bm k}}{\varepsilon_{n{\bm k}}}+
\tau_n\frac{H_{\bm k} ^{*}}{\Delta_{\bm k} }+\tau_n \frac{H_{\bm k} ^{**}}{\Delta_{\bm k} \varepsilon_{n{\bm k}}}\right),
\end{equation}
with 
\begin{align}
H_{\bm k} ^*&=(\bm h^{\bm B}_+\cdot \bm h_-)\tau_z +g_{\bm k} (\bm h^{\bm B}_+\cdot \bm \sigma) \tau_x,\notag\\
H_{\bm k} ^{**}&=(\bm h^{\bm B}_+\cdot \bm h_-)
[(\bm h_-\cdot\bm \sigma)+(\bm h^{\bm B}_+\cdot\bm \sigma)\tau_z ]+g_{\bm k}[g_{\bm k}(\bm h^{\bm B}_+\cdot\bm \sigma)+
|\bm h^{\bm B}_+|^2 \tau_x+(\bm h^{\bm B}_+\times \bm h_-) \cdot\bm \sigma  \tau_y].\notag
\end{align}
The sublattice-resolved local density of states (LDOS) is determined from the above eigenprojectors, and by further defining the sublattice projectors $P_{\alpha}=|\bm r_\alpha\rangle \langle \bm r_\alpha|=\frac{1}{2}(1+s_{\alpha}\tau_z)$ with $s_{A}=+$ for the $A$ sublattice and $s_{B}=-$ for the $B$ sublattice,
\begin{align}
%\begin{array}{ll}
\rho_\alpha(\varepsilon,\bm B)
&=\sum_{n} \langle
%A_{\bm k \sigma \tau}(\bm r,\varepsilon)
|\langle {\bm r}_\alpha |n\bm k \rangle|^2\delta(\varepsilon -\varepsilon_{n{\bm k}})
\rangle_{\bm k},
\notag\\
&=\sum_{n} \langle \textrm{Tr}\{ P_{\alpha} P_{n\bm k}\} \delta(\varepsilon -\varepsilon_{n{\bm k}}) \rangle_{\bm k},\notag\\
&=\sum_{n} \langle \frac{1}{2}(1+s_{\alpha}\tau_nv_{n\bm k})\delta(\varepsilon -\varepsilon_{n{\bm k}})
\rangle_{\bm k}
,
%\end{array}
\end{align}
with
\begin{align}
v_{n\bm k}(\bm B)=\frac{\bm h^{\bm B}_+\cdot \bm h_-}{\Delta_{\bm k}}.
\end{align}
The Brillouin zone average is denoted by $\langle \cdots \rangle_{\bm k}=\int d^d\bm k/(2\pi)^d(\cdots)$.
The above LDOS expression is valid at any order in the external magnetic field $\bm B$. 
The zero-field LDOS is obtained by the substitution $\bm h^{\bm B}_+ \rightarrow \bm h_+$.
%in all previous expressions; it rewrites
%\begin{align}
%\rho(\bm r,\varepsilon,\bm B)
%=\sum_{\sigma \tau} \int \frac{d^d\bm k}{(2\pi)^d}
%\frac{1}{2}[1+s_{\bm r}\gamma_{\bm k \sigma \tau}]\delta(\varepsilon -%\varepsilon_{\bm k \sigma \tau}),
%\end{align}
%with 
%$\gamma_{\bm k \sigma \tau}=\frac{\tau}{Y_{\bm k}}[(\bm m_{+}\cdot \bm %m_-)+\frac{f_{\bm k} }{\epsilon_{\bm k \sigma\tau}}|\bm m_{+}|^2]$.
Keeping terms to linear order in the magnetic field, one can identify the diagonal and the Berry-like contributions to LDOS,
\begin{align}
\label{eq:ldos_app}
\rho_\alpha(\varepsilon,\bm B)
=&\sum_{n} 
\langle
\frac{1}{2}[1+s_{\alpha}(\tau_n v_{n\bm k}\big|_{\bm B=0}-{\bm \Omega}_{n\bm k}\cdot \bm B)]
\delta(\varepsilon -\varepsilon_{n\bm k}\big|_{\bm B=0}-{\bm m}_{n\bm k}^d\cdot \bm B)
\rangle_{\bm k}
,
\end{align}
where from now on we work with zero-field functions $v_{n\bm k} = v_{n\bm k}(\bm B=0)$, $\varepsilon_{n\bm k}=\varepsilon_{n\bm k}(\bm B=0)$, $\Delta_{\bm k}=\Delta_{\bm k}(\bm B=0)$, etc.
The intra- and interband contributions in Eq.~\eqref{eq:ldos_app} are
\begin{align}
{\bm m}_{n\bm k}^{\rm d}&=\frac{1}{\varepsilon_{\bm k \sigma \tau}}[(1 + \tau_n u_{\bm k}) \ {\bm h_+} +
\tau_n v_{\bm k}  \ \bm h_-)],\notag\\
%{\Delta_{\bm k}}[g^2_{\bm k} {\bm h_+} +
%(\bm h_+\cdot \bm h_-)\bm h_-)],\notag\\
{\bm \Omega}_{n\bm k}&=-\tau_n\frac{ u_{\bm k}}{\Delta_{\bm k}^2}[(\bm h_+\cdot \bm h_+)\bm h_--(\bm h_+\cdot \bm h_-)\bm h_+],\notag\\
&=\tau_n \frac{u_{\bm k}}{\Delta_{\bm k}^2}(\bm h_- \times \bm h_+)\times \bm h_+\label{eq:omega_sq},
\end{align}
with
\begin{equation}
u_{\bm k}=\frac{g^2_{\bm k}}{\Delta_{\bm k}}.
\end{equation}
%where $\bm h_{\bm k}=\frac{1}{Y_{\bm k}}( g^2_{\bm k} {\bm h_+} +
%(\bm m_+\cdot \bm m_-)\bm m_-)$.
The expression of ${\bm m}_{n\bm k}^{\rm d}$ can also be recovered from the usual definition ${\bm m}^{\rm d}_{n\bm k}=\textrm{Tr} \{ P_{n\bm k}\bm \sigma\}$. 
Qualitatively, in linear order in the magnetic field, the effective spin magnetic moment ${\bm m}_{n\bm k}^{\rm d}$ essentially measures the field-induced shift of the band energy, whereas the effective spin Berry curvature ${\bm \Omega}_{n\bm k}$ measures the effective field-induced deformation of the Bloch states.
These expressions determine the local thermodynamic spin magnetization,
\begin{align}
\widetilde{\bm M}^{\alpha}_{\rm spin}=-\sum_{n} \int \frac{d^d\bm k}{(2\pi)^d} 
\big[f(\varepsilon_{n\bm k})
{\bm m}^{{\rm d},\alpha}_{n\bm k}
+F(\varepsilon_{n\bm k}){\bm \Omega}_{n\bm k}^{\alpha}\big].
\end{align}
The expressions for the diagonal and interband contributions are
\begin{align}
{\bm m}^{{\rm d},\alpha}_{n\bm k}&=\frac{1}{2}(1+s_{\alpha}\tau_nv_{n\bm k}){\bm m}_{n\bm k}^{\rm d},\label{eq:mkst_intra} \\
{\bm \Omega}^{\alpha}_{n\bm k}&=\frac{s_{\alpha}}{2}{\bm \Omega}_{n\bm k}.\label{eq:omega_r_sq}
\end{align}
%${\bm m}_{\bm k \sigma \tau}(\bm r)=\frac{1}{2}(1+s_{\bm r}\gamma_{\bm k \sigma \tau})
%{\bm m}_{\bm k \sigma \tau}$ and
%${\bm \Omega}_{\bm k \sigma\tau}(\bm r)=\frac{s_{\bm r}}{2}{\bm \Omega}_{\bm k \sigma\tau}$. 
%These identities can also be obtained from their definitions (\ref{mspin2}).
For comparison, the conventional formula Eq.~(1) from the main text yields the local spin magnetization 
\begin{align}
\bm M_{\rm spin}^\alpha
=-\sum_{n} \int \frac{d^d\bm k}{(2\pi)^d}
f(\varepsilon_{n\bm k})
\bm  m_{n\bm k}^\alpha,
\end{align}
with
\begin{align}
\bm  m_{n\bm k}^\alpha
&=\textrm{Tr} \{ P_{n\bm k} P_{\alpha}\bm \sigma \}\\
&=\frac{1}{2}[{\bm m}_{n\bm k}^{\rm d}
+\frac{s_{\alpha}}{\varepsilon_{n\bm k}} (\bm h_- +\tau_n v_{\bm k}\bm h_+)].\notag
\end{align}
The quantity $\bm  m_{n\bm k}^\alpha$ is further decomposed into a sum of intraband Eq.~\eqref{eq:mkst_intra} and interband contributions, so that the interband part reads
\begin{align}
%{\bm m}_{\bm k \sigma \tau}^{\textrm{d}}(\bm r)&={\bm m}_{\bm k \sigma \tau}(\bm r)=\frac{1}{2}(1+s_{\bm r}\gamma_{\bm k \sigma \tau}){\bm m}_{\bm k \sigma \tau}\notag\\
{\bm m}^{\textrm{od},\alpha}_{n\bm k}&=
\bm  m_{n\bm k}^\alpha-{\bm m}_{n\bm k}^{\textrm{d},\alpha},\notag\\
&=\frac{s_{\alpha}}{2}\frac{u_{\bm k}}{\varepsilon_{n \bm k}\Delta_{\bm k}}(\bm h_+ \times \bm h_-)\times \bm h_+.
%&=\frac{s_{\alpha}}{2}\frac{g^2_{\bm k}}{\varepsilon_{\bm k \sigma \tau}\Delta_{\bm k}^2}(\bm h_+ \times \bm h_-)\times \bm h_+.
\label{eq:mintra_sq}
\end{align}
Consequently, the thermodynamic theory and the conventional one yield different predictions for  the interband local spin magnetization as seen from the expressions for $\bm\Omega_{n\bm k}^\alpha$~\eqref{eq:omega_sq} and \eqref{eq:omega_r_sq} versus ${\bm m}^{\textrm{od},\alpha}_{n\bm k}$~\eqref{eq:mintra_sq}.
The above formulas determine the integrand in the local magnetization $\bm M_{\rm spin}$ and $\widetilde{\bm M}_{\rm spin}$.
Therefore, obtaining the final results presented in Fig.~1 in the main text required only a numerical integration over the Brillouin zone.

\begin{figure*}
    \centering
    \includegraphics[width=\textwidth]{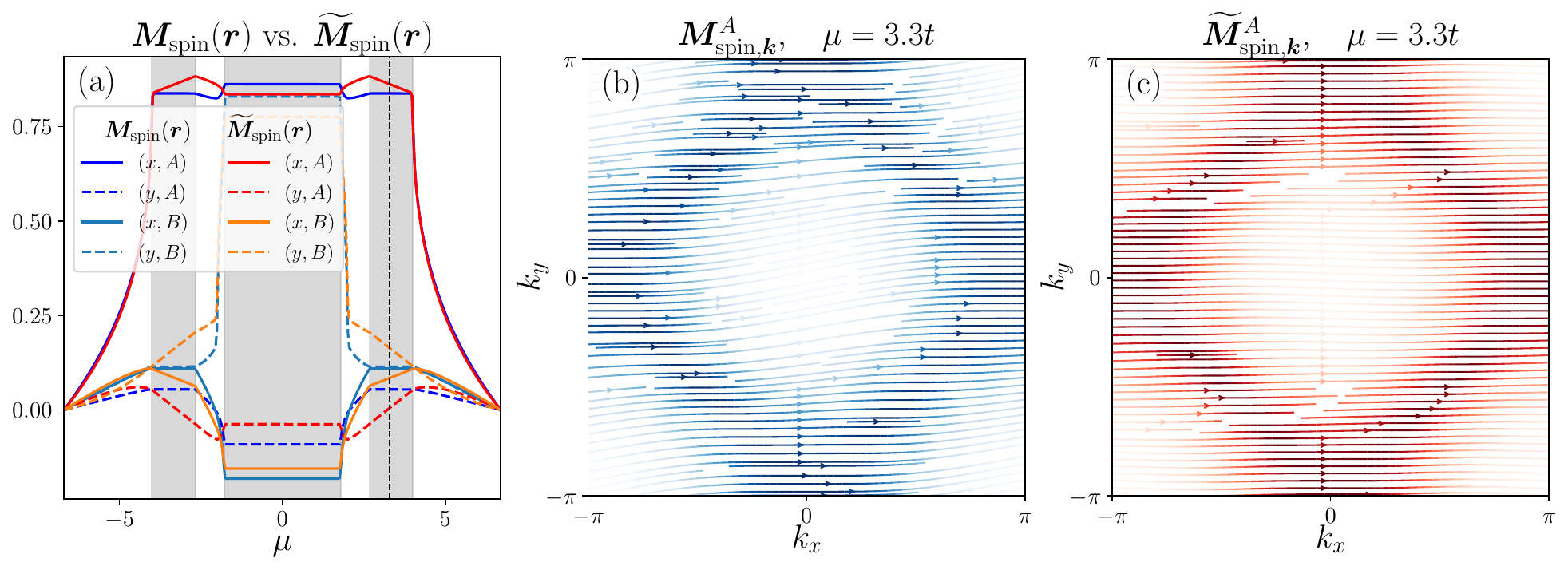}
    \caption{Comparison between the total local spin magnetization in the conventional theory [$M_{\rm spin}(\bm r)$] and the thermodynamic one [$\widetilde{M}_{\rm spin}(\bm r)$] for the same parameters as in the Fig.~1 from the main text, $\bm h_A=(4,0,0)t$, $\bm h_B=(0,2,0)t$, and temperature $T=0.01t$. 
    (a) Shows the dependence on the chemical potential for all nonzero magnetization components ($x$ and $y$) and both sublattices ($A$ and $B$). 
    Panels (b) and (c) show the momentum-resolved local spin magnetization from Eq.~\eqref{eq:mom_res} in the conventional theory and the thermodynamic one, respectively, at chemical potential $\mu=3.3t$ inside the gap, marked by the black dashed line in (a).}
    \label{fig:total_magn}
\end{figure*}

In order to complete the results shown in Fig.~1 from the main text, which capture only the interband distribution, we show here in Fig.~\ref{fig:total_magn}(a) the total local spin magnetization after adding the intraband contributions as well.
Here again, the linear dependence of magnetization on the chemical potential inside the gap, inherited from the interband components, is clearly visible.
Moreover, Figs.~\ref{fig:total_magn}(b) and (c) show the distribution of local magnetization in momentum space near a gap center, where the momentum-resolved local magnetization is defined as
\begin{align}\label{eq:mom_res}
\bm M_{{\rm spin},{\bm k}}^\alpha &= - \sum_{n} f(\varepsilon_{n\bm k})m^{\alpha}_{n\bm k},\\
\widetilde{\bm M}_{{\rm spin},{\bm k}}^\alpha & = - \sum_{n} [f(\varepsilon_{n\bm k})m^{{\rm d},\alpha}_{n\bm k}
+F(\varepsilon_{n\bm k})\Omega_{n\bm k}^\alpha].\notag
\end{align}
In contrast to the total (momentum-integrated) local spin magnetization, the momentum-resolved local spin magnetization in the two theories shows only small quantitative differences. 

\subsection{Effects of an alternating onsite potential}
In the main text, we have considered the strong-coupling regime with exchange fields of amplitude larger than the hopping strength. 
At weaker exchange fields, the side gaps in the spectrum close up, and the difference between the two theories is less pronounced, showing only quantitative differences in the local spin magnetization, but the same flat magnetization plateau in the single remaining zero energy gap.
In order to reproduce the distinctive signature in the thermodynamic theory, i.e.,~the linear variation of the local spin magnetization in the gap, we find that adding an alternating on-site potential that breaks the sublattice symmetry is sufficient.

\begin{figure}[t]
    \centering
    \includegraphics[width=\textwidth]{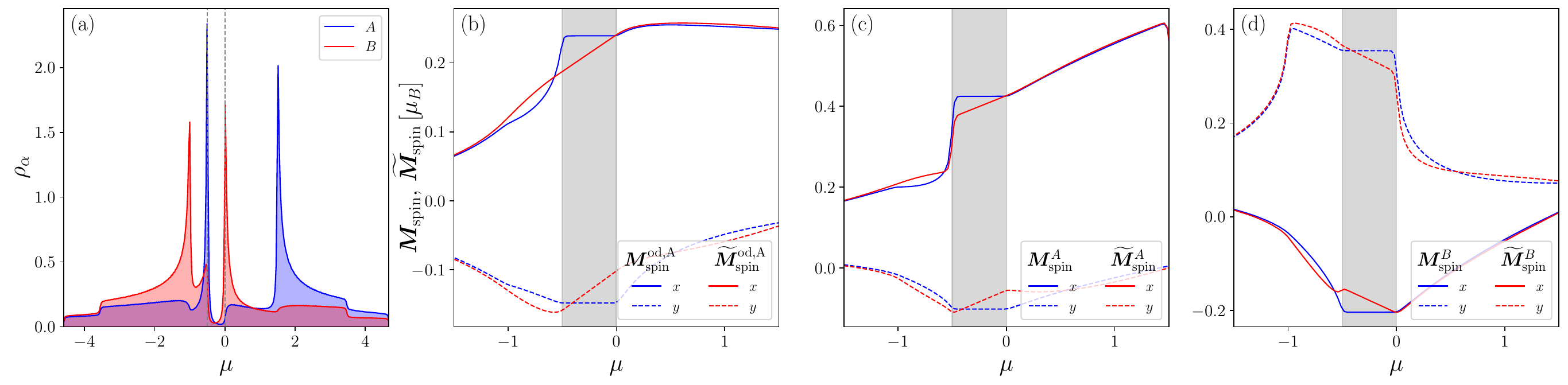}
    \caption{Effect of an alternating potential on (a) the local density of states $\rho_\alpha$ on sites $\alpha$, $A$ (blue) and $B$ (red), with dashed lines marking the spectral gap.
    Panels (b), (c), and (d) zoom near the gap (marked by the gray band) to show different components of the local spin magnetization in the conventional theory $\bm M_{\rm spin}$ against the thermodynamic one $\widetilde{\bm M}_{\rm spin}$.
    Panel (b) represents the off-diagonal magnetization on $A$ sites, while (c) and (d), the total magnetization on $A$ and $B$ sites, respectively.
    The exchange fields are $\bm h_A=(1,0,0)t$ and $\bm h_B=(0,0.5,0)t$, with an alternating potential $V=0.5t$, and temperature $T=0.01t$.}
    \label{fig:square_alt}
\end{figure}

The Hamiltonian of Eq.~\eqref{eq:ham_sq_sm} is supplemented by the onsite alternating potential $V$,
\begin{equation}
H_{\bm k} \to H_{\bm k} + V\tau_z.
\end{equation}
The model is no longer amenable to a compact analytical solution, and therefore the local spin magnetization in the two theories is determined numerically.
The exchange fields have amplitude on the order of the hopping strength, $\bm h_A=(1,0,0)t$ and $\bm h_B=(0,0.5,0)t$, with $V=0.5t$.
The resulting model has a single gap in the spectrum, reflected in the local density of states in Fig.~\ref{fig:square_alt}(a).
The two interband contributions, $\widetilde{\bm M}^{\rm od}_{\rm spin}$ and $\bm M^{\rm od}_{\rm spin}$, are shown in Figs.~\ref{fig:square_alt}(b) for sites $A$, while the corresponding results for sites $B$ have opposite in sign.
The total magnetization for both sites $A$ and $B$, with non-equal intraband contributions, on the two sublattices, is shown in Figs.~\ref{fig:square_alt}(c) and (d), respectively.
Figs.~\ref{fig:square_alt}(b)--(d) clearly show the linear variation of the local spin magnetization in the gap for the thermodynamic theory, in contrast to the flat plateaus predicted by the conventional theory.

\section{Magnetic dimer: A minimal model for non-collinear magnetism}
\label{sec:dimer}

\subsection{Model and derivation of local spin magnetization}
\begin{figure}[t]
\begin{minipage}[t]{0.35\textwidth}
\vspace{18px}
\centering
\includegraphics[width=\textwidth]{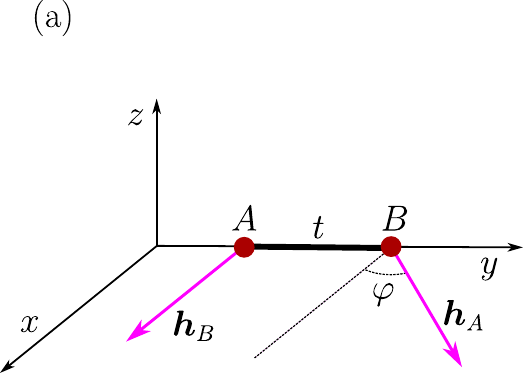}
\end{minipage}
\begin{minipage}[t]{0.45\textwidth}
\vspace{0pt}
\centering
\includegraphics[width=\textwidth]{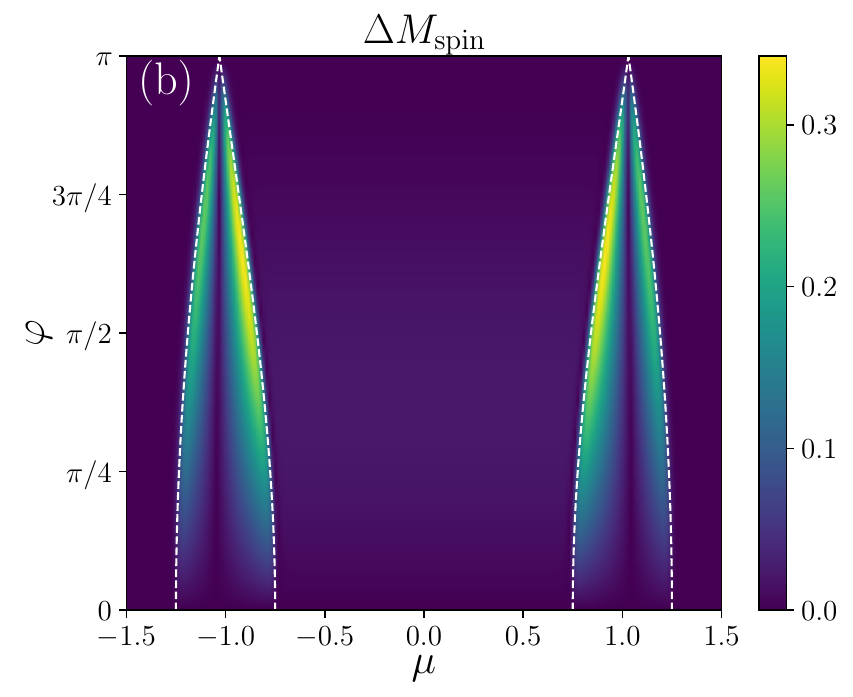}
\end{minipage}
\caption{
(a) Sketch of the magnetic dimer~\eqref{eq:mag_dimer} with intersite hopping $t$ and exchange fields $\bm h_A$ and $\bm h_B$ canted by an angle $\varphi$.
(b) Scalar difference $\Delta M_{\rm spin}$~\eqref{eq:scalar_diff} between the local spin magnetization predicted in the conventional and thermodynamic theory as a function of chemical potential $\mu$ and canting angle $\varphi$ in units of $\mu_B$. 
The energy levels are denoted by white dashed lines.
The parameters in dimensionless units read $M=1$, $t=1/4$, and temperature $T=0.01$.
}
\label{fig:dimer_diff}
\end{figure}

A minimal model for the non-collinear magnetic order that sheds light on the difference between the conventional and thermodynamic theories is that of a two-site dimer with exchange fields.
The dimer model sketched in Fig.~\ref{fig:dimer_diff}(a) is a zero-dimensional analog of the square-lattice model where the momentum dependence is removed.
The setup is sufficiently simple to allow for a fully analytical treatment, while still showcasing a redistribution of local spin magnetization between the two sites of the dimer in the thermodynamic theory, in contrast to the conventional one.
Several of the calculations and results presented for the square lattice model can be directly transposed to the dimer case, but are repeated here for completeness.
Working in the zero-field limit, the second-quantized Hamiltonian reads
\begin{align}\label{eq:mag_dimer}
\mathcal H = c_A^\dagger \bm h_A\cdot \bm \sigma \, c_A + c_B^\dagger \bm h_B\cdot \bm \sigma \, c_B+ t(c_A^\dagger c_B+c_B^\dagger c_A),
\end{align}
with $c_{\alpha}^\dag=(c_{\alpha,\uparrow}^\dag, c_{\alpha,\downarrow}^\dag)$, a vector of creation operators for site $\alpha=A,B$, and $t>0$ the spin-independent hopping amplitude.
The local exchange fields $\bm h_A$ and $\bm h_B$ are chosen without loss of generality to lie in the $x$-$y$ plane, and may be of unequal magnitude $M_A$ and $M_B$,
\begin{equation}
\bm h_A=M_A(1,0,0), \qquad \bm h_B=M_B(\cos\varphi,\sin\varphi,0).
\end{equation}
Collinear magnetic order corresponds to canting angle $\varphi=0$ or $\pi$, whereas $0<\varphi<\pi$ describes a non-collinear dimer.
The first-quantized Hamiltonian $H$ is a $4\times4$ matrix $\mathcal H = C^\dag H C$, with $C^\dag = (c_{A,\uparrow}^\dag, c_{A,\downarrow}^\dag, c_{B,\uparrow}^\dag, c_{B,\downarrow}^\dag)$ and
\begin{equation}
H = \bm h_+\cdot \bm \sigma + \bm h_-\cdot \bm \sigma\tau_z + t\tau_x,
\end{equation}
with $\bm h_\pm = \frac12(\bm h_A \pm \bm h_B)$.

By denoting 
\begin{equation}
M_\pm = \frac{M_A \pm M_B}{2},
\end{equation}
the eigenstates of the Hamiltonian are given by
\begin{equation}
\varepsilon_{\sigma\tau} = \sigma\sqrt{t^2 + M_+^2 + M_-^2 + 2\tau \Delta},
\end{equation}
and 
\begin{align}
\Delta &= \sqrt{t^2\bm h_+^2  + (\bm h_+\cdot \bm h_-)^2},\\
%&= \frac12\sqrt{
%t^2(M_A^2 + M_B^2 + 2M_AM_B\cos\varphi) + \frac14(M_A^2 - M_B^2)^2
%}\\\notag
&=\sqrt{
M^2_-(t^2+M_+^2)+t^2(M_+^2-M_-^2)\cos(\varphi/2)^2
}.
\end{align}

% In the site-pseudospin basis, this Hamiltonian takes the compact form
% \begin{align}
% \mathcal H = C^\dag H C, \qquad C^\dag = (c_{A,\uparrow}^\dag, c_{A,\downarrow}^\dag, c_{B,\uparrow}^\dag, c_{B,\downarrow}^\dag),
% \end{align}
% with $c_{x,\sigma}^\dag$ the creation operator of an electron with spin $\sigma=\uparrow,\downarrow$ on site $x=A,B$.
% The Hamiltonian matrix $H$ is 
% \begin{align}
% H = m
% \end{align}

The projectors onto the eigenstate $|\sigma\tau\rangle$ are $P_{\sigma \tau}=|\sigma \tau\rangle \langle \sigma \tau|$, and similar to the lattice results~\eqref{eq:projectors} read
\begin{align}\label{eq:projectors_dimer}
P_{\sigma \tau}=\frac{1}{4}\left(1+\frac{H}{\varepsilon_{\sigma\tau}}+
\tau\frac{H ^{*}}{\Delta}+\tau \frac{H ^{**}}{\Delta \varepsilon_{\sigma\tau}}\right),
\end{align}
with 
\begin{align}
H ^*&=(\bm h_+\cdot \bm h_-)\tau_z + t(\bm h_+\cdot \bm \sigma) \tau_x,\notag\\
H ^{**}&=(\bm h_+\cdot \bm h_-)
[(\bm h_-\cdot\bm \sigma)+(\bm h_{+}\cdot\bm \sigma)\tau_z ]+t[t(\bm h_{+}\cdot\bm \sigma)+
|\bm h_{+}|^2 \tau_x+(\bm h_+\times \bm h_- )\cdot\bm \sigma  \tau_y].\notag
\end{align}
Similarly, we introduce the orbital projectors $P_\alpha = |\alpha\rangle \langle \alpha| = \frac{1}{2}(1 + s_\alpha \tau_z)$, with $\alpha=A,B$, and $s_{A/B}=\pm1$.

The local spin magnetization on each orbital follows after a straightforward calculation,
\begin{align}
\bm m^{{\rm d}, \alpha}_{\sigma\tau} &= \text{Tr}\{P_{\sigma\tau}P_\alpha\}\text{Tr}\{P_{\sigma\tau}\bm \sigma\}\notag\\
&=\frac{1}{2}(1+s_\alpha\tau u)\bm m_{\sigma\tau}^{\rm d},
\end{align}
with
\begin{align}
u&=\frac{\bm h_+\cdot \bm h_-}{\Delta} = \frac{M_+ M_-}{\Delta},\\
\bm m_{\sigma\tau}^{\rm d}&=\frac{1}{\varepsilon_{\sigma\tau}}\left[\bm h_+
+\frac{\tau}{\Delta}\left(t^2\bm h_++(\bm h_+\cdot \bm h_-)\bm h_-\right)\right]\\
&=\frac{1}{\epsilon_{\sigma\tau}}(w_+M_A+w_-M_B\cos\varphi, w_-M_B\sin\varphi, 0),
\end{align}
with
\begin{equation}
w_\pm = \frac{1}{2\Delta}(\Delta\pm \tau M_+M_-+\tau t^2).
\end{equation}

The Berry curvature-like contribution to the local magnetization is determined by
\begin{align}\label{eq:omega_st}
\bm\Omega_{\sigma\tau}^\alpha &= \frac{s_\alpha}{2}\frac{\tau t^2}{\Delta^3}(\bm h_-\times \bm h_+)\times \bm h_+,\\
&=\frac{s_\alpha\tau t^2 M_AM_B\sin\varphi}{8\Delta^3}\bm v_\varphi,
\end{align}
with the vector controlling the off-diagonal response
\begin{equation}
\bm v_\varphi = (-M_B\sin\varphi,\,M_A+M_B\cos\varphi,\,0),
\end{equation}
As expected, for collinear magnetism ($\varphi=0$ or $\pi$) the Berry curvature-like contribution generally vanishes, whereas it is finite for non-collinear textures.

By comparison, the conventional off-diagonal contribution reads
\begin{align}
\bm m^{{\rm od},\alpha}_{\sigma\tau} &= \frac{s_\alpha t^2}{2\varepsilon_{\sigma\tau}\Delta^2}(\bm h_+\times \bm h_-)\times \bm h_+,\\
&=-\frac{s_\alpha t^2 M_AM_B\sin\varphi}{8\varepsilon_{\sigma\tau}\Delta^2}\bm v_\varphi.
\end{align}
As in the thermodynamic theory, the off-diagonal contribution is generically vanishing for collinear order $(\varphi=0 \text{ or } \pi)$.

Note that both off-diagonal contributions have different sign on the two orbitals, $s_{A/B}=\pm1$, and therefore the total off-diagonal contribution to the dimer's magnetization is zero.
Nevertheless, local probes could reveal the different distribution of spin magnetization on different sites.

\subsection{Equal magnitude exchange fields}
To gain more insight into the differences between the two theories, it is instructive to focus on the case of equal-magnitude exchange fields, 
\begin{equation}
M \equiv M_A=M_B,
\end{equation}
where the expressions simplify significantly.
Here, $M_+=M$ and $M_-=0$, so that the wave-function has equal weight on both orbitals ($u\equiv M_+M_-/\Delta=0$).
The energy spectrum simplifies to
\begin{equation}
\varepsilon_{\sigma\tau} = \sigma\sqrt{t^2 + M^2 + 2\tau tM\cos(\varphi/2)},
\end{equation}
with $\varphi$ restricted to the interval $[0,\pi]$.
Note that the antiferromagnetic configuration with $\varphi=\pi$ is special since each energy eigenvalue becomes doubly degenerate, $\varepsilon_{\sigma\tau}(\varphi=\pi) = \sigma\sqrt{t^2 + M^2}$, and the projectors are defined only on the two-dimensional degenerate subspace. This case will be discussed separately below.

The diagonal contribution to local spin magnetization now reads
\begin{align}
\bm m^{\rm d,\alpha}_{\sigma\tau} 
&=\frac{|\bm h_+|+\tau t}{2\varepsilon_{\sigma\tau}}\frac{\bm h_+}{|\bm h_+|},\\
&=\sigma
\frac{M\cos\frac{\varphi}{2}+\tau t}{2\sqrt{M^2+t^2+2\tau tM\cos\frac{\varphi}{2}}}\notag
(\cos\frac{\varphi}{2},\sin\frac{\varphi}{2},0).
%= \frac{M}{4\varepsilon_{\sigma\tau}}
%(1 + \frac{\tau t^2}{Y})(1+\cos\varphi, \sin\varphi, 0),
\end{align}
% with
% \begin{align}
%     \bm m_+&=\frac{M}{2}(1+\cos\varphi,\sin\varphi,0),\\
% \varepsilon_{\sigma\tau} &= \sigma\sqrt{t^2 + M^2 + 2\tau Y},\quad
% Y = t|\bm m_+|,
% \end{align}
with $\varphi$ restricted to the interval $[0,\pi]$.
The intraband local magnetization vector $\bm m_{\sigma}^{\rm d,\alpha}$ is therefore parallel to the sum of local exchange fields $\bm h_+=(\bm h_A+\bm h_B)/2$.

The off-diagonal contribution in the thermodynamic theory is now
\begin{align}
\label{eq:omega_st_eq}
\bm \Omega_{\sigma\tau}^\alpha &= -\frac{s_\alpha\tau}{2t}\frac{\bm h_-}{|\bm h_+|}\\
&=\frac{s_\alpha\tau \sin(\frac{\varphi}{2})}{2t}
(-\tan\frac{\varphi}{2},1,0).\notag
\end{align}
The counterpart off-diagonal contribution in the conventional theory reads
\begin{align}
\label{eq:mintra_sq_eq}
\bm m^{\rm od,\alpha}_{\sigma\tau} &= \frac{s_\alpha}{2\varepsilon_{\sigma\tau}}\bm h_-,\\
&= \frac{s_\alpha\sigma M \sin\frac{\varphi}{2}}{2\sqrt{M^2+t^2+2\tau tM\cos\frac{\varphi}{2}}}
(\sin\frac{\varphi}{2},-\cos\frac{\varphi}{2},0).\notag
\end{align}
Both contributions are thus in the direction of the difference between the local exchange fields, $\bm h_-$.

\subsubsection{The antiferromagnetic limit}
The limit of collinear antiferromagnetic order of the above expressions is not well defined, since $\bm h_+\to 0$, as well as $\tau\to0$. 
As each pair of energy levels becomes degenerate, the off-diagonal contribution between is converted to a diagonal contribution, i.e., within the degenerate subspace.
Up to an overall spin rotation, the Hamiltonian in the antiferromagnetic limit $\varphi=\pi$ reads
\begin{equation}
H = M\sigma_x\tau_z + t\tau_x,
\end{equation}
with energy eigenvalues 
\begin{equation}
\varepsilon_\sigma=\sigma\sqrt{t^2 + M^2},\quad \sigma=\pm1.
\end{equation}
The projectors onto the degenerate subspaces indexed by $\sigma$ are
\begin{align}
P_{\sigma} = \frac{1}{2}\left(1+\frac{H}{\varepsilon_{\sigma}}\right).
\end{align}
Note that the eigenstates of the Hamiltonian for the collinear case are also eigenstates of spin, $\sigma_x$.
Within each degenerate subspace, let $\eta$ denote a spin index for $\pm$ eigenstates of $\sigma_x$. 
The energy eigenvalues do not depend on $\eta$.
Within the degenerate subspace, the projectors onto the spin eigenstates are
\begin{align}
P_{\eta} = \frac{1}{2}(1+\eta\sigma_x).
\end{align}
Then, the total local spin magnetization in the antiferromagnetic limit reads
\begin{equation}
\bm m_{\sigma}^\alpha = \text{Tr}\{P_{\sigma}P_\alpha\bm \sigma\} = \frac{\sigma s_\alpha M}{\sqrt{M^2+t^2}}\hat{\bm x},\quad
\hat{\bm x} = (1,0,0).
\end{equation}
The diagonal contribution to the local magnetization reads
\begin{align}
\bm m^{\rm d,\alpha}_{\sigma} &= \sum_\eta \text{Tr}\{P_{\sigma}P_\eta P_\alpha\}\text{Tr}\{P_{\sigma}P_\eta\bm \sigma\},\\
&=\frac{\sigma s_\alpha M}{\sqrt{M^2+t^2}}\hat{\bm x}.
\end{align}
As expected, the entire off-diagonal contribution is converted to a diagonal one within the degenerate subspace, such that the off-diagonal contribution vanishes,
\begin{equation}
\bm m^{\rm od,\alpha}_{\sigma} = \bm 0.
\end{equation}

The Berry-like contribution to the local magnetization also needs to be carefully evaluated in the antiferromagnetic limit, although taking the limit $\varphi \to \pi$ in Eq.~\eqref{eq:omega_st} yields the correct vanishing result.
Using the Green's function formalism for the degenerate subspaces,
\begin{equation}
G(\varepsilon) = \sum_{\sigma} \frac{1+H/\varepsilon_{\sigma}}{\varepsilon - \varepsilon_{\sigma}+i0^+},
\end{equation}
and Eqs.~(9) in the main text, it follows that the off-diagonal contribution to the local magnetization in the antiferromagnetic limit is
\begin{equation}
\widetilde{\bm M}_{\rm spin}^{\rm od,\alpha} 
=-\sum_{\sigma}\int d\varepsilon F(\varepsilon) 
\frac{\text{Im}}{\pi}
\text{Tr}\bigg\{
\frac12(1+s_\alpha\tau_z)
\frac{1+H/\varepsilon_{\sigma}}{\varepsilon - \varepsilon_{\sigma}+i0^+}
\bm \sigma
\frac{1-H/\varepsilon_{\sigma}}{\varepsilon + \varepsilon_{\sigma}+i0^+}
\bigg\}.
\end{equation} 
However, the integrand vanishes due to trace properties
\begin{equation}
\text{Tr}\bigg\{
(1+s_\alpha\tau_z)
\big(1+\frac{H}{\varepsilon_{\sigma}}\big)
\bm \sigma
\big(1-\frac{H}{\varepsilon_{\sigma}}\big)
\bigg\}=0,
\end{equation}
leading to vanishing off-diagonal or the Berry-like contributions to the local magnetization in the antiferromagnetic limit,
\begin{equation}
\bm \Omega_{\sigma}^\alpha(\varphi=\pi) = \bm 0.
\end{equation}
Only the diagonal contribution survives, and it is identical to the conventional theory, as expected.
With the antiferromagnetic limit resolved, we can now compare the two theories in the general non-collinear case.

\subsection{Local spin magnetization near zero temperature}
We have seen that in both conventional and thermodynamic theories, the off-diagonal correction is purely local: it redistributes the magnetization between the two sites but cancels when adding the total magnetization from both sites.
To provide an quantitative illustration of the difference between the two theories, we compute the local magnetization as a function of the chemical potential $\mu$ in the zero temperature limit.

Continuing with the case of equal-magnitude exchange fields, let us focus on the off-diagonal terms which carry the difference between the two theories. 
Using Eqs.~\eqref{eq:omega_st_eq} and \eqref{eq:mintra_sq_eq}, the off-diagonal contributions are
\begin{align}
\bm M_{\rm spin}^{\rm od,\alpha}&=-\sum_{\sigma,\tau}f(\varepsilon_{\sigma\tau})\frac{s_\alpha}{2\varepsilon_{\sigma\tau}}\bm h_-,\\
\widetilde{\bm M}_{\rm spin}^{\rm od,\alpha} & = \sum_{\sigma,\tau}F(\varepsilon_{\sigma\tau})\frac{s_\alpha\tau}{2t|\bm h_+|}\bm h_-.
\end{align}
At zero temperature the Fermi factors reduce to
\begin{align}
f(\varepsilon)\simeq\Theta(\mu-\varepsilon),
\quad
F(\varepsilon)\simeq(\mu-\varepsilon)\Theta(\mu-\varepsilon),
\end{align}
with $\Theta(x)$ the Heaviside step function.

Let us now compute the interband local magnetization as a function of the chemical potential $\mu$. The spectrum consists of two particle-hole symmetric pairs of levels, $E_+ > E_-> -E_- > -E_+$, with
\begin{align}
E_\pm \equiv \varepsilon_{\sigma=+,\tau} = \sqrt{t^2 + M^2 + 2t\tau M\cos\frac{\varphi}{2}},
\end{align}
and becomes degenerate at the antiferromagnetic point $\varphi=\pi$ where $E_+=E_-$.
Other degeneracies occur at specific parameters, e.g., $M=t$, which can be avoided in the following discussion, by assuming either strong coupling $t\gg M$ or weak coupling $t\ll M$.

The off-diagonal local magnetization has a step-wise dependence on $\mu$ as the different levels are crossed,
\begin{align}
\bm M_{\rm spin}^{\rm od,\alpha}&=\frac{s_\alpha\bm h_-}{2}
\begin{cases}
\frac{1}{E_+}, & -E_+<\mu<-E_-,\\
\frac{E_++E_-}{E_+E_-},    & -E_-<\mu<E_-,\\
\frac{1}{E_+}, & E_-<\mu<E_+,\\
0, & \text{otherwise}.
\end{cases}
\end{align}
In the antiferromagnetic limit $\varphi \to \pi$, the level $E_+$ and $E_-$ become degenerate $E=E_+=E_-$, and as shown before it vanishes in both theories.

The thermodynamic local magnetization behaves quite differently due to the linear dependence on $\mu$,
\begin{equation}
\widetilde{\bm M}_{\rm spin}^{\rm od,\alpha}=
\frac{s_\alpha\bm h_-}{2t|\bm h_+|}
\begin{cases}
E_++\mu, & -E_+<\mu<-E_-,\\
E_+-E_-, & -E_-<\mu<E_-,\\
E_+-\mu, & E_-<\mu<E_+,\\
0, & \text{otherwise}.
\end{cases}
\end{equation}
At $\varphi=\pi$, the off-diagonal contribution vanishes.

The linear dependence on $\mu$ at odd filling for the levels, explains also the Fig.~1 in the main text, where the same phenomenology is observed in the square lattice model, in the outer gaps of the spectrum, where an odd number of bands is filled.

To quantify the difference between the two theories, let us define the scalar difference between the two off-diagonal contributions 
\begin{equation}
\label{eq:scalar_diff}
\Delta M_{\rm spin}(\mu)
=\left|
\widetilde{\bm M}_{\rm spin}^{\rm od,\alpha}(\mu)-
\bm M_{\rm spin}^{\rm od,\alpha}(\mu)
\right|
\end{equation}
A plot of the quantity is shown in Fig.~\ref{fig:dimer_diff}(b) as a function of the chemical potential $\mu$ and the canting angle $\varphi$ close to zero temperature, in the strong-coupling regime $M>t$.
As shown by analytics, the maximum difference is obtained when an odd number of levels is occupied, between the first and second level, and between third and fourth level, when the thermodynamic interband contribution is linear in $\mu$.
In the collinear limits $\varphi=0$ and $\pi$, the difference vanishes, and is maximal for intermediate canting angles $\varphi\sim\pi/2$.

\section{Magnetic dice lattice}\label{sec:dice}
\begin{figure*}[t]
\includegraphics[width=0.20\textwidth]{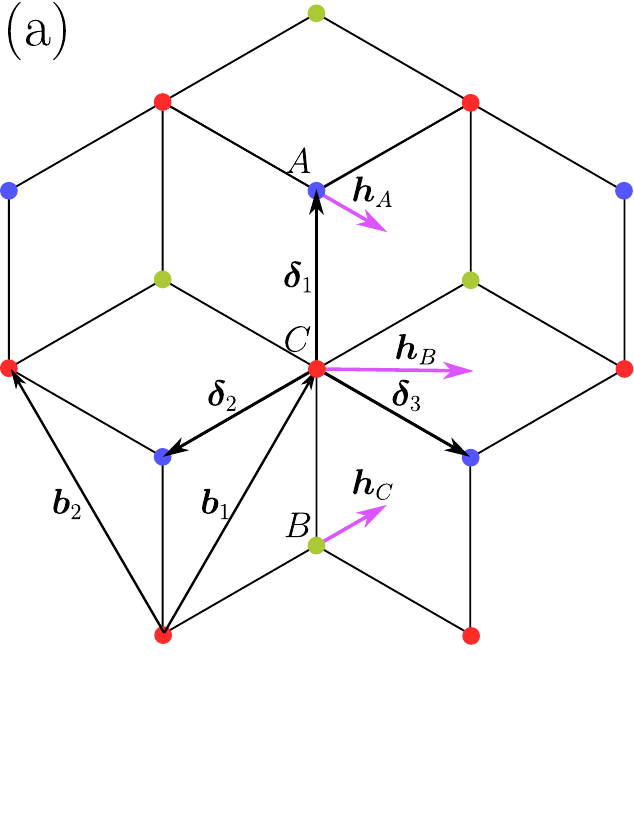}
\includegraphics[width=0.35\textwidth]{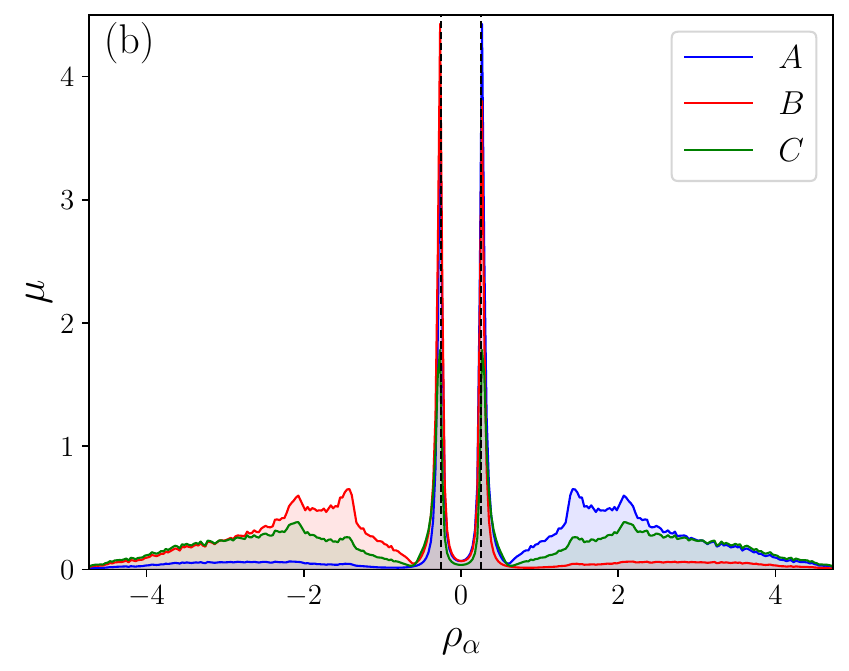}
\includegraphics[width=\textwidth]{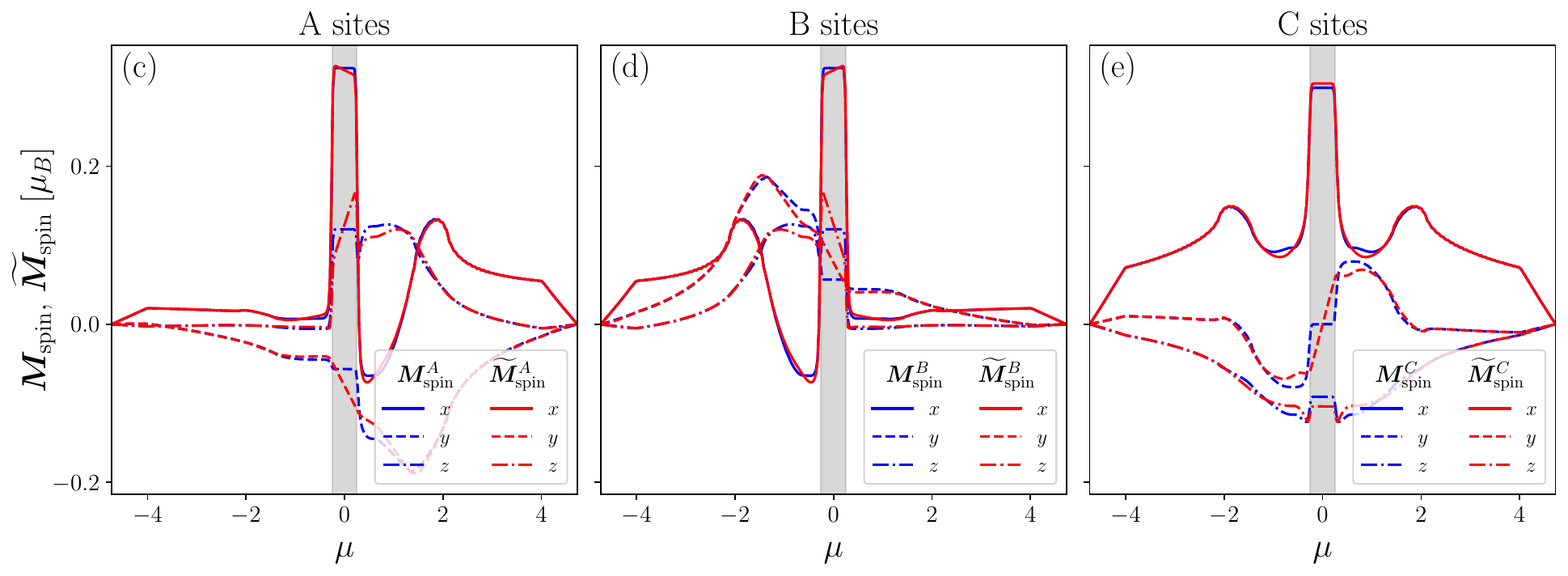}
\caption{(a) Schematic representation of the dice lattice with primitive vectors $\bm b_1$ and $\bm b_2$, nearest-neighbor vectors $\bm \delta_1$, $\bm \delta_2$, and $\bm \delta_3$, and (magenta) the projection of exchange fields $\bm h_{A,B,C}$ in the lattice plane.   
(b) Local density of states at the three sites. 
Panels (c), (d), and (e) show the total local spin magnetization in the conventional theory (blue) $\bm M_\text{spin}$ against the thermodynamic one (red) $\widetilde{\bm M}_\text{spin}$ at site A, B, and C, respectively. 
The legend indicates which of the magnetization components are measured, $x$, $y$, or $z$.
The onsite potentials are $V_A=t$, $V_B=-t$, and $V_C=0$ and the temperature is $T=0.01t$.
}
\label{fig:dice}
\end{figure*}
The dice lattice tight-binding model is described by a chiral symmetric Hamiltonian with three sites per unit cell, featuring a three-band spectrum, with a characteristic Dirac cone and a middle flat band.
The model is a playground where the differences between the conventional and thermodynamic theories of local spin magnetization can be clearly illustrated.
Here we consider a spinful version of the model with onsite potentials that gap the Dirac spectrum.
The idea explored in this section is to add local exchange fields with magnitude weaker than hopping, with in- and out-of-plane components, in order to show the generality of the results from the main text.

The dice lattice Hamiltonian reads
\begin{equation}
H = C^\dag_{\bm k} H_{\bm k} C_{\bm k}, \quad C^\dag_{\bm k} = (c^\dag_{\uparrow,A}, c^\dag_{\downarrow,A}, c^\dag_{\uparrow,B}, c^\dag_{\downarrow,B}, c^\dag_{\uparrow,C}, c^\dag_{\downarrow,C}),
\end{equation}
with Bloch Hamiltonian
\begin{align}
H_{\bm k} &= 
\begin{pmatrix}
V_A + \bm h_A\cdot \bm \sigma & 0 & g(\bm k) \\
0 & V_B + \bm h_B\cdot \bm \sigma & g(\bm k)^* \\
g(\bm k)^* & g(\bm k) & V_C + \bm h_C \cdot\bm \sigma
\end{pmatrix},\notag\\
g(\bm k) &= -{}t(e^{i\bm k\cdot \bm \delta_1}+e^{i\bm k\cdot \bm \delta_2}
+e^{i\bm k\cdot \bm \delta_3}).
\end{align}
The nearest-neighbor vectors are $\bm \delta_1=(0,1,0)$, $\bm \delta_2=\frac12(-\sqrt 3, -1,0)$, $\bm \delta_3=\frac12(\sqrt 3, -1,0)$, while the exchange fields are
$\bm h_A = \frac t 4(\frac{\sqrt 3}{2},-1, \frac12)$, $\bm h_B = \frac t 4(\frac{\sqrt 3}{2},1,\frac12)$, $\bm h_C = \frac t 4(2,0,-1)$.
Fig.~\ref{fig:dice}(a) shows a dice lattice patch annotated with the projection of $\bm h_\alpha$ vectors on the lattice plane, with the $\bm \delta_i$ vectors, and the primitive vectors $\bm b_1=\frac 1 2(\sqrt 3, 3, 0)$ and $\bm b_2=\frac 1 2(-\sqrt 3, 3, 0)$.
The exchange fields lift the spin-degeneracy of the dice-lattice flat band, and a gap emerges at zero energy. 
Flat bands are still present at the gap edges, resulting in singularities in the local density of states [Fig.~\ref{fig:dice}(b)].
The local spin magnetization is computed in both conventional and thermodynamic theories, and the results are shown in Figs.~\ref{fig:dice}(c--e) for sites $A$, $B$, and $C$, respectively.
As in the previous models, the two theories predict different local spin magnetization, with the thermodynamic theory showing a linear variation of the magnetization in the gap, while the conventional theory predicts a flat plateau.

\section{Skyrmion texture and spin-orbit coupling on the hexagonal lattice}\label{sec:soc_skyrmion}

In the following, we illustrate the difference between the two theories in a nontrivial magnetic texture with spin-orbit coupling, an alternating onsite potential, and a skyrmion field on the hexagonal lattice.
The tight-binding Hamiltonian for the system writes
\begin{equation}
\label{eq:H_skyrmion_2}
H = \sum_{i,\alpha}[V_\alpha c_{\alpha,i}^\dag c_{\alpha,i} + \bm c^\dag_{\alpha,i} \bm h(\bm r_{i,\alpha})\cdot\bm \sigma c_{\alpha,i}]
+t\sum_{\langle i,j\rangle} (c_{A,i}^\dag c_{B,j} + \text{H.c.})
-i\sum_{\langle\!\langle i,j\rangle\!\rangle,\alpha} \big[\lambda_\alpha c_{\alpha,i}^\dag\bm e_{ij} \cdot\bm \sigma c_{\alpha,j} - \text{H.c.}\big],
\end{equation}
where $i$ runs over the unit cells and $\alpha$ runs over the sublattices $A$ and $B$. 
The sum over $\langle\dots\rangle$ runs over the nearest-neighbor sites, while $\langle\!\langle\dots\rangle\!\rangle$ runs over next-nearest-neighbor sites.
The Rashba spin-orbit coupling (SOC) term is parametrized with the amplitude $\lambda_{A/B}=\pm\lambda$ and
the alternating potential is $V_{A/B}=\pm V$.
The spin indices for the creation (and annihilation) operators are implied with $c_{j,\alpha}^\dag=(c^\dag_{j,\alpha,\uparrow}, c^\dag_{j,\alpha,\downarrow})$.
The next-nearest-neighbor vectors are $\bm e_{ij}=\bm r_j-\bm r_i$, where $i, j$ denote the $A$ sites in nearest-neighboring cells. The skyrmion field is defined as
\begin{equation}
\bm h(\bm r) = M\bigg(\frac {x}{r}\sin\theta_r, \frac {y}{r}\sin\theta_r, \cos\theta_r \bigg),
\end{equation}
with $\bm r=(x,y)$ running over the lattice sites, $r=|\bm r|$, $\theta_r=\pi r/R$.

A strong alternating onsite potential $V$ opens a gap, while the joint action of spin-orbit coupling and the skyrmion field produce a linear dependence of local spin magnetization inside the gap.

\begin{figure}[t]
    \centering
    \includegraphics[width=\textwidth]{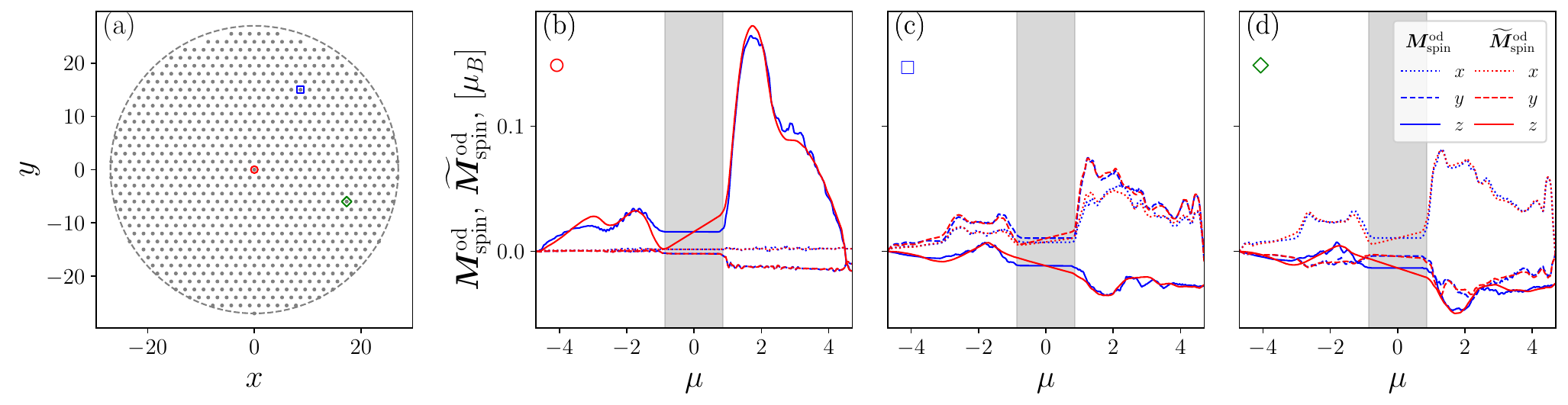}
    \caption{(a) Sketch of the hexagonal lattice circumscribed by a circle of radius $R=27a$ hosting the skyrmion texture.
    (b), (c), and (d) show the off-diagonal local spin magnetization in the conventional (blue lines) and the thermodynamic (red lines) theories for three $A$ sites indicated in panel (a) by symbols $\circ$, $\smallsquare$, and $\diamond$, respectively.
    The other parameters are: skyrmion field amplitude $M=0.2t$, spin-orbit coupling strength $\lambda=0.2t$, onsite potential $V=1.5t$, and temperature $T=0.01t$.
    The gray regions mark the spectral gaps seen in the local density of states.
    The legend is shared by (b), (c), and (d) and indicates the three spatial components of the local magnetization.
    }
    \label{fig:soc_skyrmion}
\end{figure}

In Fig.~\ref{fig:soc_skyrmion}, we present the results at three different points in the lattice for the off-diagonal component of the local spin magnetization in the conventional and the thermodynamic theory.
Outside the gap, $\bm M^{\rm od}_{\rm spin}$ follows closely $\widetilde{\bm M}^{\rm od}_{\rm spin}$, although some quantitative differences exist.
Notably, inside the gap, the components of $\widetilde{\bm M}^{\rm od}_{\rm spin}$ show a linear dependence on the chemical potential, in contrast to the flat plateaus in the conventional theory.

\bibliography{references}